\documentclass[aps,prl,twocolumn,groupedaddress,showpacs]{revtex4}
\usepackage{bm,graphicx}

\begin{document}

\title{Dynamics of cholesteric structures in an electric field}

\author{O.~S. Tarasov$^{1,2}$, A.~P. Krekhov$^{1,2}$, and L.~Kramer$^1$}
\affiliation{
$^1$Physikalisches Institut, Universit\"at Bayreuth,
D-95440 Bayreuth, Germany\\
$^2$Institute of Molecule and Crystal Physics,
Russian Academy of Sciences, 450075 Ufa, Russia}

\date{March 25, 2003}

\begin{abstract}
Motivated by Lehmann-like rotation phenomena in cholesteric drops we
study the transverse drift of two types of cholesteric fingers, which 
form rotating spirals in thin layers of cholesteric liquid crystal
in an {\em ac} or {\em dc} electric field.
We show that electrohydrodynamic effects induced by Carr-Helfrich charge
separation or flexoelectric charge generation can describe the drift of
cholesteric fingers.
We argue that the observed Lehmann-like phenomena can be understood
on the same basis.
\end{abstract}

\pacs{61.30.Gd, 61.30.Jf, 47.20.Ky}

\maketitle

\begin{figure}[ht]
\includegraphics[width=7cm,height=3cm]{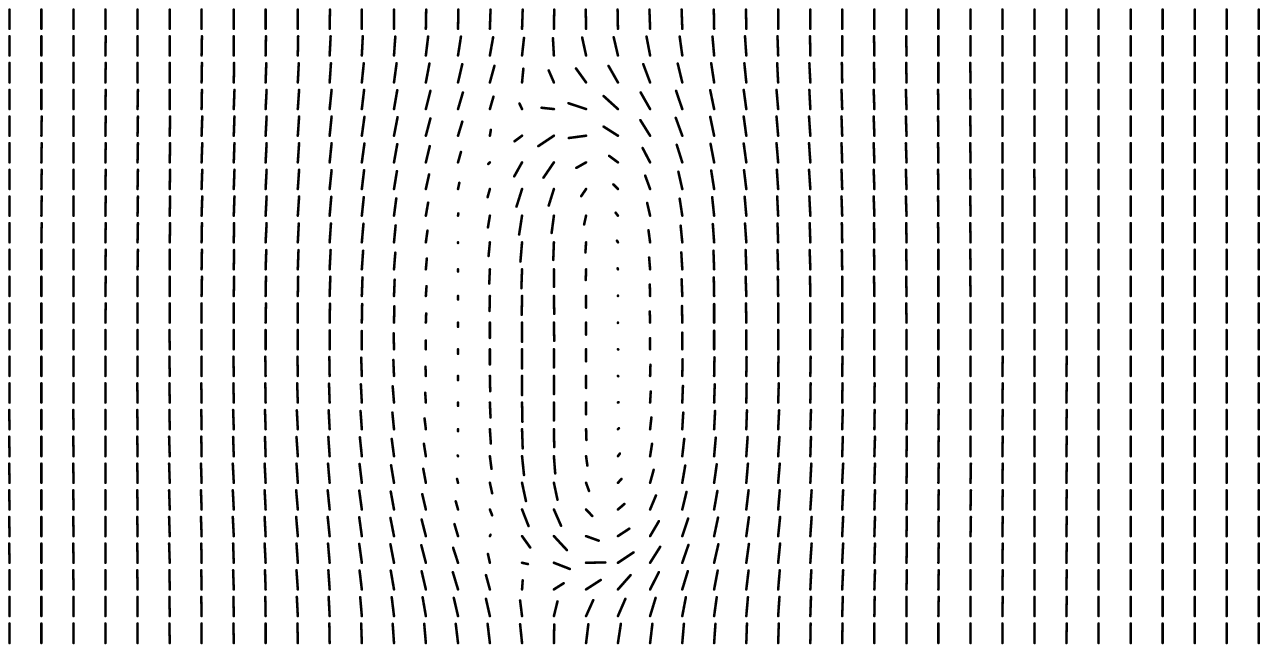}\\
\includegraphics[width=7cm,height=3cm]{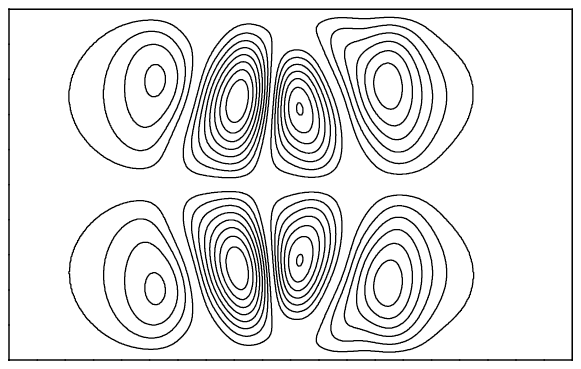}
\caption{Director profile (top) and induced
velocity profile (stream lines) (bottom) of CF2. $U=1.9$V, $\omega
\tau_q=1$, 5CB material parameters \protect\cite{5cb}.}
\label{CF2_profiles}
\end{figure}
\begin{figure}[ht]
\includegraphics[width=7cm,height=3cm]{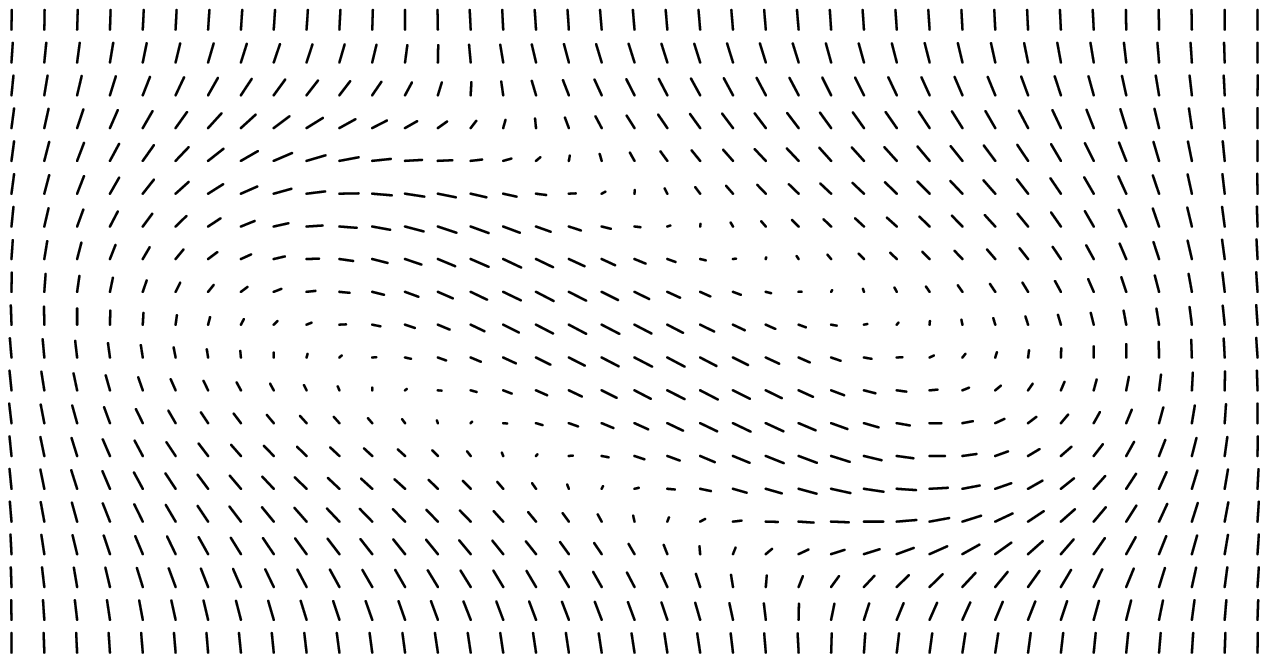}\\
\includegraphics[width=7cm,height=3cm]{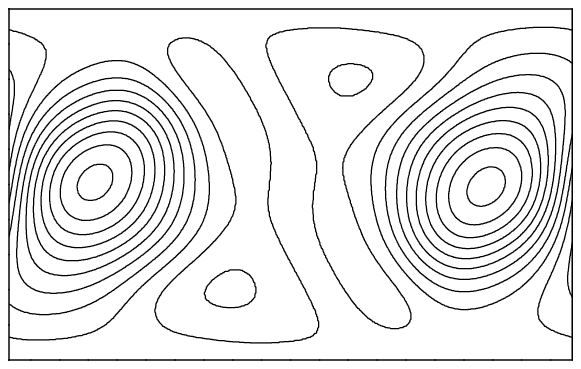}
\caption{
Director profile (top) and induced
velocity profile (stream lines) (bottom) of CF1. $U=0.2$V,
$L=L_F=1.95$, MBBA material parameters \protect\cite{mbba}.}
\label{CF1_profiles}
\end{figure}

%
One of the most interesting manifestations of macroscopic chirality are 
cholesteric liquid crystals.
In cholesterics, as in nematic liquid crystals, there is long-range 
orientational order of the elongated molecules along a local axis 
described by the director  $\bm{n}$.
Whereas in nematics the elastic forces (or torques) tend to establish a 
uniform orientation of $\bm{n}$, the chiral molecules in cholesterics 
lead in equilibrium to a helical arrangement with $\bm{n}$ 
perpendicular to the helix axis.
Choosing the axis along $z$, the structure with pitch $p_0$ is given by
$\bm{n}=(\cos{\varphi(z)},\sin{\varphi(z)},0)$, $\varphi=q_0 z$, 
$q_0=2 \pi/p_0$.
The helical symmetry leads to interesting dynamical effects.
Particularly intriguing is the Lehmann rotation of the director 
structure in a cholesteric droplet heated from below,
described in 1900 \cite{Leh1900,degennes,chandrabook}.
It has not been observed again, which is usually attributed to the 
influence of surface anchoring.
However, the electric analog where the temperature gradient is 
replaced by a {\em dc} electric field $\bm{E}$ (``electromechanical 
effect'') has been observed \cite{madpra87}.
Traditionally, the explanation is based on phenomenological hydrodynamic
considerations, which by symmetry allow for an additional 
dissipative dynamic coupling between the director and 
electric field (``electromechanical coupling'')
\cite{degennes,chandrabook}.
Strictly speaking, here ``hydrodynamic'' means that spatial modulations
must be slow on  the scale of the pitch which is not the case in
\cite{madpra87}.
In fact, in the experiments the director structure inside the droplet 
was not a homogeneous helix, but included substantial spay-bend 
distortions and even defects.
Another weakness of the approach is that no underlying mechanism has 
been identified that would, at least in principle, allow to determine 
the coefficients involved.
We have investigate for the first time a driving mechanism for 
chirality-related dynamical phenomena, involving well-established 
electro-hydrodynamic (EHD) effects.
Actually, director rotation is not the best choice to study such
phenomena, since special precautions are needed to avoid surface 
anchoring \cite{madpra87}.
Another effect is related to the intrinsic length scale defined by
helical symmetry.
In a confined geometry this can lead to the spontaneous formation
of spatial structures (``cholesteric fingers'') lacking certain
reflection  symmetries, which under nonequilibrium conditions results
in drift of the structures.
This is not affected by surface anchoring.
We will show that this drift can be explained by EHD effects and,
at the end, return to the rotation phenomena.
Cholesterics placed between two plane parallel electrodes (separation 
$d$), providing strong homeotropic (perpendicular to the electrodes) 
anchoring of $\bm{n}$, can experience unwinding of the helix due to 
the orienting effect of the electric field and of the boundaries.
The unwinding transition, which is typically discontinuous, occurs 
when a combination of the confinement ratio $C=d/p_0$ and the 
electric field strength (or applied voltage $U$) reaches a critical
value \cite{obpPR2000}.
Thus, there exists a line in the $(U,C)$ plane where the two phases 
coexist.
Near this line one finds the cholesteric fingers (CFs): elongated 
structures that are localized or arranged periodically.
At least four types of CFs were observed \cite{bpoPRE1998}.
The director configuration of CF of the first type (CF1) is invariant
with respect to a $\pi$-rotation about the finger axis 
\cite{prearrJP1976}, whereas CF2 has a mirror symmetry with respect to 
the midplane of the cell \cite{gilgilPRL1998,bpoPRE1998}, see 
Figs. \ref{CF2_profiles}, \ref{CF1_profiles} (top).
The difference in structure of CF1 and CF2 manifests itself in the 
dynamics.
In a {\em dc} field both fingers are observed to drift perpendicular to
their axis  \cite{gilgilLC1994,gilthiJP1997}, whereas in an {\em ac} 
field only CF2 drifts \cite{gilgilLC1994,ropJP1994,gilthiJP1997}.
Since CFs also grow along their axis one observes the formation of CF 
spirals.
Most measurements of the transverse drift of CFs are based on the 
analysis of spiral dynamics.
As the motor of the CF1 drift only the electromechanical coupling has 
been proposed \cite{gilthiJP1997}.
The magnitude of the drift velocity estimated with the coupling 
coefficient taken from droplet rotation-type experiments  turns out 
to be at least one order higher than observed \cite{gilthiJP1997}.
To explain the drift of CF2 in an {\em ac} field, several models were 
proposed \cite{gilgilPRL1998}, including  electromechanical coupling.
However, these models fail to describe the recently observed falloff of
the drift velocity $V_{\perp}$ when the frequency $f=\omega/2\pi$ of 
the applied electric field approaches the inverse charge relaxation time
$\tau_q$ \cite{bpoPRE1999}.
In the experiments the conductivity, and thereby $\tau_q$, was varied by
using different concentrations of ionic dopant.
%

%
The mechanism we propose is based on flow induced either by charge
separation through anisotropic conductivity, i.e., the Carr-Helfrich 
effect, (CF2 under {\em ac} driving) or by flexoelectric charge 
generation (CF1 under {\em dc} driving).
We use the standard set of nematodynamic equations for the director
$\bm{n}$, the velocity $\bm{v}$ (Navier-Stokes equation) in the Stokes
approximation (neglect of inertial terms), and the electric field 
$\bm{E}$ \cite{degennes,chandrabook}.
As usual the electric properties of the material are described by a 
dielectric permittivity tensor
$\epsilon_{i j}=\epsilon_\perp \delta_{i j} + \epsilon_a n_i n_j$,
($\epsilon_\perp$, $\epsilon_\parallel=\epsilon_\perp+\epsilon_a$ 
are the permittivities perpendicular and parallel to the director,
respectively), the analogous conductivity tensor
$\sigma_{i j}=\sigma_\perp \delta_{i j} + \sigma_a n_i n_j$,
and a flexopolarisation
$\bm{P}^{fl}=e_1\bm{n}(\nabla\cdot\bm{n}) + 
e_3 (\bm{n}\cdot\nabla)\bm{n}$.
We introduce dimensionless variables
$t = \tau_d \tilde{t}$,
$\bm{r} = d \tilde{\bm{r}}$,
$\bm{E} = (U_0/d) \tilde{\bm{E}}$,
$\bm{P}^{fl} = (\epsilon_0 U_0/d)\tilde{\bm{P}}^{fl}$,
and charge density
${\rho}_{el}=(\epsilon_0 U_0/d^2)\tilde{\rho}_{el}$.
Here $\tau_d = \gamma_1 d^2/K_{33}$ is the relevant director relaxation
time ($\gamma_1=$ rotational viscosity, $K_{33}=$ bend elastic
constant), $U_0=\sqrt{K_{33}/\epsilon_0}$ characterizes the typical
voltage ($U_0\approx 1 $V for the materials used). 
We choose the $z$ axis perpendicular to the bounding electrodes and 
allow for a drift of the structure with velocity $V_\perp$ in the $x$ 
direction, transverse to the finger's long axis  by replacing
$\partial_t \rightarrow \partial_t - V_\perp \partial_x$.
Then the equations can be written as (tildes are omitted, 
$\bm{V}_\perp=V_\perp {\hat x}$)
\begin{eqnarray}
&&[\partial_t + (\bm{v}-\bm{V}_\perp)\cdot\nabla
+\gamma'_2 \underline{\underline{\delta}}^\perp \underline{\underline{A}}
-\bm{\Omega}\times] \bm{n}  =
-\underline{\underline{\delta}}^\perp \bm{h}^r ,
\label{tb_eqn}\\
&&p_{, i} - T^v_{j i, j} + h^v_k n_{k, i} =
\rho_{el} E_i - V_{\perp} S_{j i, j} ,
\label{ns_eqn}\\
&&\frac{\tau_q}{\tau_d}
[\partial_t + (\bm{v}-\bm{V}_\perp)\cdot\nabla] \rho_{el} + \rho_{el}
=
\nonumber \\
&& \;\;\;\;\;\;\;\;\;\;\;\;\;\;\;\;\;\;\;\;\;\;\;\;\;\;\;\;
\nabla\cdot[ - \epsilon_\perp \xi_H(\bm{n}\cdot\bm{E})\bm{n} 
+ \bm{P}^{fl} ] .
\label{cc_eqn}
\end{eqnarray}
supplemented by the incompressibility condition $\nabla\cdot\bm{v}=0$,
the Poisson equation $\rho_{el}=\nabla\cdot[ \epsilon_\perp \bm{E}
+\epsilon_a (\bm{n}\cdot\bm{E})\bm{n}+\bm{P}^{fl}]$, which was
already used in the charge conservation equation (\ref{cc_eqn}),
the electrostatic condition $\nabla\times\bm{E}=0$,
and the director normalization $\bm{n}^2=1$.
The notation $f_{, i}=\partial f/\partial x_i$ is used throughout.
The generation of space charges is characterized by the
Helfrich parameter
$\xi_H=\sigma_a/\sigma_\perp - \epsilon_a/\epsilon_\perp$
in Eq.  (\ref{cc_eqn}).
The director equation (\ref{tb_eqn}) couples to the flow field through 
the local fluid rotation
$\bm{\Omega}=(\nabla\times\bm{v})/2$ 
and the hydrodynamic strain tensor 
$A_{i j}=(v_{i, j} + v_{j, i})/2$ 
with 
$\gamma'_2=\gamma_2/\gamma_1$ 
and the projection tensor
$\delta_{i j}^\perp=\delta_{i j} - n_i n_j$.
Coupling to the elastic and electric torques is through
$h_i^r = \delta F/\delta n_i$
with the free energy density
$F = \frac{1}{2} k_1 (\nabla\cdot\bm{n})^2 +
\frac{1}{2} k_2 (2\pi C + \bm{n}\cdot(\nabla\times\bm{n}))^2
+ \frac{1}{2} k_3 (\bm{n}\times(\nabla\times\bm{n}))^2 -
\frac{1}{2} \epsilon_a (\bm{n}\cdot\bm{E})^2 -
\bm{P}^{fl}\cdot\bm{E}$.
Here $k_i=K_{ii}/K_{33}$ with the elastic constants $K_{ii}$.
In Eq. (\ref{ns_eqn}) the Stokes approximation is justified since the 
processes of interest are controlled by $\tau_d \sim 1$ s and the
charge relaxation time 
$\tau_q=\epsilon_0\epsilon_\perp/\sigma_\perp \sim 10^{-3}$ s,
which are much larger than the viscous relaxation time
$\tau_v=\rho_m d^2/\gamma_1 \sim 10^{-6}$ s ($\rho_m=$ mass density).
The elastic part of the stress tensor (Ericksen tensor) has been 
eliminated and the pressure redefined:
$p=p_0+F$ \cite{Leslie_ALC4_1_1979}.
The  viscous  stress tensor is
$T_{i j}^v = \alpha'_1 n_i n_j n_k n_m A_{k m} +
          \alpha'_2 n_i N_j +
          \alpha'_3 n_j N_i
      + \alpha'_4 A_{i j} +
          \alpha'_5 n_i n_k A_{k j} +
          \alpha'_6 n_j n_k A_{k i}$,
where $\alpha'_i=\alpha_i/\gamma_1$ with the Leslie viscosity 
coefficients $\alpha_i$.
Moreover 
$\bm{N}=(\partial_t+\bm{v}\cdot\nabla)\bm{n} - \bm{\Omega}\times\bm{n}$
and 
$h_k^v=N_k +\gamma'_2 n_j A_{j k}$.
The bulk force $\rho_{el}\bm{E}$ is the Coulomb force and 
$S_{ij}=\alpha'_2 n_i \partial_x n_j+\alpha'_3 n_j \partial_x n_i$.
%

%
We solve Eqs.(\ref{tb_eqn})-(\ref{cc_eqn}) in a perturbative way:
$\bm{n} = \bm{n}_0 + \bm{n}_1 + \dots$,
$\bm{E} = \bm{E}_0 + \bm{E}_1 + \dots$,
$\bm{v} =  \bm{v}_1 + \dots$,  $V_\perp=V_{\perp 1}+\dots$,
$\rho_{el} =  \rho_{el 1} + \dots$.
At lowest order the electric charge is neglected.
Therefore no bulk force arises in Eq.(\ref{ns_eqn}) and thus
$\bm{v}_0\equiv 0$.
$\bm{n}_0$ and $\bm{E}_0$ are obtained  from
\begin{eqnarray}
\label{n0_eqn}
\underline{\underline{\delta_0}}^\perp \bm{h}_0^r = 0 , \;\;
\nabla\cdot[
\epsilon_\perp \bm{E}_0 +
\epsilon_a (\bm{n}_0\cdot\bm{E}_0)\bm{n}_0] = 0 .
\label{E0_eqn}
\end{eqnarray}
We write
$\bm{E}_0=E_0(\bm{\hat z} - \nabla\phi_0)\cos(\omega\tau_d t)$
with $E_0$ the applied electric field and $\phi_0$ the induced 
potential.
Previous results show that the director configurations obtained from 
Eqs.(\ref{n0_eqn}) and the stability diagrams are in good agreement
with the experiments \cite{obpPR2000,bpoPRE1998,bpoPRE1999}.
We have solved Eqs.(\ref{n0_eqn}) in two dimensions (infinite extension
of the CF in the $y$ direction) by a relaxation method.
At first order Eq.(\ref{cc_eqn}) gives
\begin{eqnarray}
\label{rho_el1}
\rho_{el 1} &=&
-\frac{\cos(\omega\tau_d t) + \omega\tau_q\sin(\omega\tau_d t)}
      {1 + \omega^2\tau_q^2}
E_0 \epsilon_\perp \xi_H
\nonumber\\
&&\times
\nabla\cdot\{ [\bm{n}_0\cdot(\bm{\hat z}-\nabla\phi_0)]\bm{n}_0 \}
\nonumber\\
&&+e_{fl}
\nabla\cdot\{ \bm{n}_0(\nabla\cdot\bm{n}_0)
+ \frac{e_3}{e_1}(\bm{n}_0\cdot\nabla)\bm{n}_0 \} ,
\end{eqnarray}
where $e_{fl}=e_{1} d/(\epsilon_0 U_0)$.
The {\em dc} case is covered by setting $\omega=0$.
Next one solves the Navier-Stokes Eq.(\ref{ns_eqn}) (together with the 
incompressibility condition) at first order.
Taking the $\bm{curl}$ eliminates the pressure and leads to an 
inhomogeneous ODE for $\bm{v}_1$ with the coefficients depending on
$\bm{n}_0$.
The form of the inhomogeneities [right-hand side of (\ref{ns_eqn})] 
give solution of the form
\begin{equation}
\bm{v}_1 = 
\frac{\epsilon_\perp \xi_H E_{eff}^2}{1+\omega^2\tau_q^2} \bm{f}_1
 + e_{fl} E_0 \bm{f}_2 
 + V_{\perp 1} \bm{f}_3 ,
\label{v1}
\end{equation}
where $E_{eff}=E_0/\sqrt{2}$ ($E_0$) for {\em ac} ({\em dc}) driving and
the second term appears only in the {\em dc} case (then also 
$\omega=0$).
For the {\em ac} case a time  average of Eq.(\ref{ns_eqn}) was taken.
The functions $\bm{f}_i$ depend only on $\bm{n}_0$ and the $\alpha'_i$
($\bm{f}_2$ depends also on the ratio of the flexocoefficients 
$e_3/e_1$).
We have determined the functions $\bm{f}_i$ [from the linearized 
Eq.(\ref{ns_eqn})] by a Galerkin expansion with trigonometric
functions (for CF1) and Hermite polynomials (for CF2) as trial
functions.
Finally, the director equation (\ref{tb_eqn}) at first order gives
\begin{eqnarray}
\label{n1_eqn}
\underline{\underline{\delta_0}}^\perp \bm{h}_1^r +
\underline{\underline{\delta_1}}^\perp \bm{h}_0^r &=&
(\bm{v}_1\cdot\nabla - V_{\perp 1} \partial_x)\bm{n}_0
\nonumber\\
&&+ \gamma'_2 \underline{\underline{\delta_0}}^\perp
\underline{\underline{A_1}} \bm{n}_0 
- \bm{\Omega}_1\times\bm{n}_0 ,
\end{eqnarray}
where $\bm{h}_1^r$, $\underline{\underline{\delta_1}}^\perp$ are
linear in $\bm{n}_1$.
The homogeneous problem
$\underline{\underline{\delta_0}}^\perp \bm{h}_1^r +
\underline{\underline{\delta_1}}^\perp \bm{h}_0^r = 0$
is solved by the translation mode $\partial \bm{n}_0/\partial x$.
The solvability condition for the inhomogeneous problem [with $\bm{v}_1$
substituted from (\ref{v1})] fixes the drift velocity $V_{\perp 1}$.
In physical units we obtain
\begin{equation}
V_{\perp 1}= \frac{d}{\gamma_1}
\frac{\epsilon_0 \epsilon_\perp \xi_H}
     {1+\omega^2\tau_q^2} E_{eff}^2 \frac{I_1}{I_0+I_3}
     + \frac{e_1 E_0}{\gamma_1} \frac{I_2}{I_0+I_3} ,
\label{vd}
\end{equation}
where 
$I_0 = \langle \bm{n}_{0, x}\cdot\bm{n}_{0, x} \rangle$,
$I_1 = \langle \bm{n}_{0, x}\cdot\bm{g}_1 \rangle$,
$I_2 = \langle \bm{n}_{0, x}\cdot\bm{g}_2 \rangle$,
$I_3 = \langle \bm{n}_{0, x}\cdot\bm{g}_3 \rangle$.
The last term appears only in the {\em dc} case.
The functions $\bm{g}_1$, $\bm{g}_2$, $\bm{g}_3$ are easily expressed 
in terms of the $\bm{f}_1$, $\bm{f}_2$, $\bm{f}_3$, respectively.
The scalar product is defined by
$\langle \bm{a}\cdot\bm{b} \rangle = \int\int(\bm{a}\cdot\bm{b})dx dz$.
%

\begin{figure}[ht]
\includegraphics[scale=0.45]{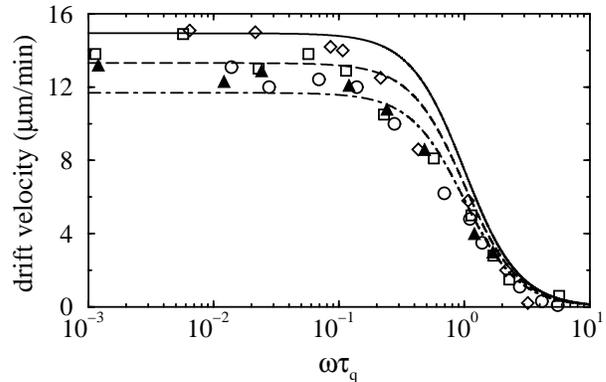}
\vspace*{-0.5cm}
\caption{
Drift velocity of CF2 versus reduced frequency: points
are experimental data from \protect\cite{bpoPRE1999},
lines are calculations from (\protect\ref{vd}):
for $\sigma_\parallel/\sigma_\perp=1.3$ (solid),
$\sigma_\parallel/\sigma_\perp=1.4$ (dashed),
and $\sigma_\parallel/\sigma_\perp=1.5$ (dash-dotted).
$C=1.77$, $d=31\,\mu$m, $U_0=1.9$ V, 5CB material parameters
\protect\cite{5cb}.}
\label{cf2comp}
\end{figure}

%
We first discuss our results on CF2 drift in an {\em ac} field.
Since in the experiments the arms of CF2 spirals are well separated 
\cite{bpoPRE1999}, we have in the computations chosen a box width
large compared to the width of the director structure (isolated
finger).
The director profile  $\bm{n}_0$ together with the stream lines of the
flow from (\ref{v1}) as computed for the  parameters of the material 5CB
(4'-n-pentyl-4-cyanobiphenyl) \cite{5cb} used in the experiments are
shown in Fig. \ref{CF2_profiles}.
In Fig. \ref{cf2comp} our results for $V_{\perp 1}$, which is of order 
of $v_1$, versus reduced frequency $\omega \tau_q$ are shown, together
with the experimental results (the different symbols relate to different
impurity concentrations between $2 \cdot 10^{-5}$ and $0.05$ wt\%)
\cite{bpoPRE1999}.
The authors of \cite{bpoPRE1999} have scaled the frequency down by a 
factor of about $1.75$ in order to account for the fact that the
charge relaxation time $\tau_q$ was measured in the isotropic phase at
temperature $T=40$ $^\circ$C (the experiments were done at
$T=30$ $^\circ$C).
The three curves correspond to $\sigma_\parallel/\sigma_\perp=1.3$,
$1.4$ and $1.5$, which agrees well with measurements in pure 5CB where
$\sigma_\parallel/\sigma_\perp=1.44$ \cite{jadz}.
$\sigma_\parallel/\sigma_\perp$ is the only parameter not given in
\cite{bpoPRE1999}.
Another experimental result born out by the model is the
approximate linear dependence of the drift velocity of CF2
as a function of applied field for samples with different
thicknesses but fixed confinement ratio $C$ \cite{bpoPRE1999}.
The CF2 appear only around an electric field such that
$(E_0 d)^2 = 
\pi^2/(\epsilon_0 \epsilon_a)
(4C^2K_{22}^2- K_{33}^2)/K_{33}$
\cite{obpPR2000}.
Using this to eliminate $d$ from (\ref{vd})
one obtains the linear dependence of the drift velocity on $E_0$.
%

%
We now turn to CF1 in a {\em dc} field.
From the symmetry of the director profile $\bm{n}_0$ one now has 
$I_1=0$, so one is left with the last term in Eq.(\ref{vd}) relating to
flexoelectric charge generation.
Unfortunately one now has to cope with various uncertainties.
First, the approximation of isolated fingers is not valid, since in the
experimental spirals neighboring fingers are packed closely
\cite{gilthiJP1997,hinkukMCLC84}.
Second, in the {\em dc} case one has to expect screening of the electric
field by Debye layers, as evidenced in \cite{hinkukMCLC84} and also
suggested by the voltage offset $\sim 2$V in the current-voltage curve
presented in \cite{gilthiJP1997} [both using the material MBBA 
(4-methoxy-benzylidene-4'-n-butylaniline)].
Thus we have solved Eqs.(\ref{n0_eqn}) on a width $L$ in the $x$ 
direction  with periodic boundary conditions.
$V_{\perp 1}$ turns out to be sensitive to $L$: it decreases with 
decreasing $L$ and even changes sign at $L\approx 1.6$.
In the absence of experimental data on the finger width we have -- on 
one hand -- minimized the free energy density with respect to $L$ for
given values of the electric field leading to the ``optimal'' box
width $L_F(U)$.
Typical values of $L_F$ (in units of $d$) are $2 \div 2.5$ for a voltage
$0.2 \div 3$V.
In Fig. \ref{CF1_profiles} an example of the calculated director profile
and the stream lines of the flow from (\ref{v1}) are shown.
\begin{figure}[ht]
\includegraphics[scale=0.45]{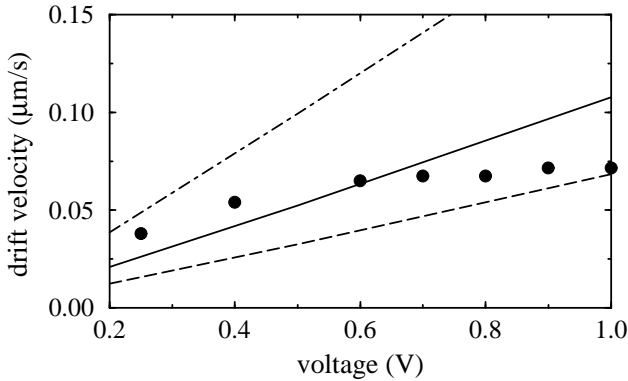}
\vspace*{-0.5cm}
\caption{
Drift velocity of CF1 versus voltage: points
are experimental data from \protect\cite{gilthiJP1997} shifted by $2$V;
lines are calculations from (\protect\ref{vd}).
Solid line: $L=L_F(U)$; dashed: $L=1.9$; dot-dashed: $L=2.1$.
Flexocoefficients $e_{1}=-1.05\cdot 10^{-11}$ C/m,
$e_{3}=-1.25\cdot 10^{-11}$ C/m.
$C=1.14$, $d=12\,\mu$m, MBBA material parameters
\protect\cite{mbba}.}
\label{cf1comp}
\end{figure}
We have also made calculations for some fixed values of $L$.
Our results for $V_{\perp 1}$ as a function of $U$, corrected by a 
screening of $2$V, are given in Fig. \ref{cf1comp}, together with the
experimental data \cite{gilthiJP1997}.
Typical values of the flexocoefficients for pure MBBA were chosen
\cite{dmdJPL82}.
%

%
In conclusion, we have developed an EHD model for the drift of CF1 and
CF2.
For CF1 in a {\em dc} electric field, flow induced by flexoelectric 
charge generation can describe the drift, but no quantitative conclusion
could be drawn.
For CF2 a Carr-Helfrich-like mechanism describes {\em quantitatively}
the drift in an {\em ac} electric field.
Moreover, our preliminary studies show that the EHD model with 
flexoelectric charge generation (as in CF1) can describe the rotation 
of cholesteric droplets in a {\em dc} electric field \cite{madpra87}, 
which has hitherto been interpreted in terms of phenomenological
electromechanical coupling.
Furthermore, it seems very likely that analogous thermo-hydrodynamic 
effects, which lead to very efficient convection phenomena in liquid
crystals \cite{thermo}, can also describe the original Lehmann
rotation \cite{Leh1900}.
This then suggests that so far there is no clear experimental 
manifestation of the (unspecified) phenomenological electro- or
thermo-mechanical coupling.
The EHD model provides a very general mechanism for forces and motion
that can be applied to other director structures, like  other types of 
cholesteric fingers (CF3 and CF4) and defects in an electric field.
%

%
We thank J. Baudry, and P. Oswald for very useful discussions on their
experiments and help with the computations of the director structures. 
Financial support by DFG Grant Kr690/14-1, RFBR Grant 02-02-17435 and
the European graduate school ``Nonequilibrium phenomena and phase 
transitions in complex systems'' funded by DFG are gratefully 
acknowledged.

%

%

\end{document}